**2015.12.03     2018.2.25:** *http://www.astro.ku.dk/~erik/xx*  inserted here, but the list stored in the Rigsarkivet in November 2016 has references to the invalid dropbox!!!

# Archives on astronomy from the 1950s

*Erik Høg     (Niels Bohr Institute, Copenhagen, Denmark)*

**Abstract:** Information on archives from the 1950s of 15 astronomical observatories is provided beginning with a list of correspondence and other information related to astronomy of the Copenhagen University Observatory in the 1950s. The Appendix contains information from the 14 other observatories about their archives from those years, most of them having no archive at all. Public links are given to most of the files. - **Print of the present list and the Danish astronomy archive itself will be placed at the *Rigsarkivet*, the Danish National Archives.**

## 1.  Introduction

Information on archives from the 1950s of 15 astronomical observatories is provided beginning with a list of correspondence and other information related to astronomy of the Copenhagen University Observatory in the 1950s. An Appendix contains information from the 14 other observatories about their archives from those years, most of them having no archive at all. Public links are given to some of the files.

It was a surprise to find that nine out of the first ten observatories inquired had no archive at all from those years, only Copenhagen had a substantial archive which is presented below. Five other observatories were then asked and it turned out that they all had an archive.

**Content:**

Section 2: The correspondence between Julie Vinter Hansen and Bengt Strömgren on 42 pages was used in writing the article, my memoirs, Høg (2015) as described in the section. This work led to location of the other correspondence registered here.

Section 3: The correspondence of Copenhagen Observatory 1947-59 is presently placed at the Kroppedal Museum. It measures in total thickness about 74 cm, with 10 pages per mm this means about 7000 letters and other correspondence of administrative and scientific character.

Sections 4 and 5: The correspondence A.1-12 with Peter Naur is related to the first *Baltic Meeting* Høg (2015a) which took place in Lund in September 1957. The letters A.13-15 are about Erik Høg.

Section 6: More on the 1950s, interviews with Peter Naur and Erik Høg.
 Appendix: "Archives of 18 observatories".

## 2.  Correspondence Julie Vinter Hansen – Bengt Strömgren 1951-58

Collected by Erik Høg in April 2015

ABSTRACT: In early 1996, the Copenhagen Observatory moved from Brorfelde and Østervold to the Rockefeller Complex at Juliane Maries Vej 30, in Copenhagen together with geophysicists and the Danish Space Research Institute. Before the move, the archives at Østervold were registered at the initiative of the director professor Henning E. Jørgensen (1938-2010) who began as a young student of astronomy in September 1956. The registrant and papers were then ready to be sent to the Science Archive in Aarhus, according to recent information from Claus Thykier (*1939), founder and leader of the Ole Rømer Museum, now named Kroppedal Museum, who supported this work of registration.

In connection with the move, copies of this correspondence, *incomplete and all in Danish*, came into my hands about 1996, I do not remember how. The originals have presently, April 2015, not been found. I have received negative answers from the archive in Aarhus and from the Rigsarkivet, the Danish National Archives. They are being searched at the Kroppedal Museum by Lene Skodborg and collaborators.

For my memoirs, I have extracted in English translation how carefully my education was discussed and arranged by Bengt Strömgren, Julie Vinter Hansen and my mentor Peter Naur. Study of the stability of the new meridian circle in Brorfelde became my task and that led me deep into astrometry.

The article Høg (2015) contains my memoirs about 1946-58. Memoirs about the years 1958-80 are available in Høg (2014). My fellow student and also astronomer Svend Laustsen has written for his children in Laustsen (2015) using the same correspondence.

The correspondence on 42 pages, B.1-42, numbered chronologically by me, has been scanned into four pdf files.

Page:

*B.1-9: Feb. 1951 – jan. 1953* http://www.astro.ku.dk/~erik/xx/CorrJulie1951.pdf

1. Julie Vinter Hansen til Bengt Strömgren  5 feb 1951

2, 3, 4-5, 6, 7, 8.  JVH til BS

9.  JVH til BS   31 jan. 1953

*B.10-19: Maj 1953 – dec. 1953* http://www.astro.ku.dk/~erik/xx/CorrJulie1953.pdf

10-11. JVH til BS   22 maj 1953

12, 13-14, 15.  JVH til BS

16-17.  BS til JVH

18-19.  JVH til BS   18 Dec 1953

B. *20-29: Jan. 1954 – marts 1956* http://www.astro.ku.dk/~erik/xx/CorrJulie1954.pdf

20-21. BS til JVH, p.1 is missing 2 Jan 1954

22, 23.  JVH til BS

24.  BS til JVH

25, 26, 27.  JVH til BS

28-29.  BS til JVH   11 marts 1956

*B.30-42: Marts 1956 – maj 1958* http://www.astro.ku.dk/~erik/xx/CorrJulie1956.pdf

30-31.  JVH til BS   27 Marts 1956

32, 33, 34, 35, 36, 37.  JVH til BS

38-39.  BS til JVH

40.  JVH til BS

41-42.  BS til JVH   8 maj 1958

## 3. Correspondence Copenhagen Observatory 1947-59

Listed by Erik Høg and Lars Occhionero on 29 April 2015.

The total thickness of about 74 cm and assuming 10 pages per mm, indicates about 7000 letters and other correspondence of administrative and scientific character.

The correspondence is placed at the Kroppedal Museum, near Copenhagen, in a box 40x40x70 cm. The box contains 15 letter box files labeled alphabetically A, B, … , V-Ø containing letters from 1947-59. The label and thickness in cm of the letters in each box are:

A 3,      B 5,       C 5,       D-E 5,     F-G 5 = 23 cm
H-J 6,    K-L 5,     M 3,       N 5,       O-P 5 = 24 cm
Q-R 7,    Sa-Sl 5,   Sm-Sø 6,   T-U 3,     V-Ø 6 = 27 cm
Total thickness = 74 cm

**The file H-J with Otto Heckmann:**
The correspondence with Otto Heckmann is contained in 13 letters between 24.04.1947 and 17.02.1959. Letter nr. 12 is from Strömgren, dated 21.08.1953, a few days before his departure to USA as director of Yerkes Observatory, and it is his last letter to OH. Letter nr. 13 is from Anders Reiz to OH. Thus, no letter about the Baltic meetings has been found among the 13.

The correspondence with Erik Høg is found on 32 pages dated from 29.07.1953 to 17.04.1959.

**Note:** The original letters of the 42 pages listed in section 2 could not be found in spite of careful search.

**Phone numbers:**
Kroppedal Museum: Kroppedals Alle 3, 2630 Taastrup.         Phone: 4330 3000
Lars Occhionero, intern fastnet 113, mobil 2624 2868
Erik Høg: 4449 2008

## 4. Correspondence with Peter Naur during 1957

*A.1-9: About the first Baltic Meeting:*
*http://www.astro.ku.dk/~erik/xx/CorrNaur1957.pdf*
1. Julie Vinter Hansen to Otto Heckmann
2. Naur to Heckmann

3. Heckmann to Naur
4. Heckmann to Gyldenkerne and Naur
5. Heckmann to Naur
6. Naur to Heckmann
7. Heckmann to Naur
8. Heckmann to Naur
9. Naur to Heckmann

## 5. Correspondence with Peter Naur during 1958

*A. 10-12: About the visit by Peter Naur to Hamburg in January 1958*
*A.13-15: About Erik Høg*
*http://www.astro.ku.dk/~erik/xx/CorrNaur1958.pdf*

10. Astronomisches Colloquium auf der Hamburger Sterwarte, Jan. und Feb. 1958
11. Naur to Heckmann after Naur's colloquium on 11 Jan. 1958
12. Naur to Dieckvoss
13. Recommendation to Høg, in Danish
14. Letter to J. Oort sending Høg's report on automatic measurement
15. Letter Høg to Naur, 3 pages in Danish

## 6. More on the 1950s

Interviews with Peter Naur and Erik Høg were recorded, in Danish, in 2009 for an historical research project. They are available with audiofile and transcribed at the Kroppedal Museum for research purposes. (File .zip of 52 MB for Naur at link in my mail to Aaserud on 28.04.2015.)

**Acknowledgements:** The author is grateful to the following persons for information and support: Anthony Brown, Lars Buus, Bengt Edvardsson, Christine Etienne, Michael Geffert, an anonymous librarian of the Paris Observatory, Wolfram Kollatschny, Jan Lub, Palle Lykke, Francois Mignard, Lars Occhionero, Javier Montojo Salazar, Gregory Shelton, Frederic Thevenin, Axel Wittmann, Norbert Zacharias and all the persons acknowledged in Høg (2015a).

# Appendix
## Archives of 15 observatories

**Abstract:** The archives from the 1950s of 15 astronomical observatories are reviewed in this report. On Copenhagen Observatory as given above, followed in this appendix by 14 other observatories. A correspondence with nine observatories between Bonn and Uppsala in early 2015 showed to my surprise that none of these observatories had an archive from those years, only Copenhagen had a substantial archive as presented above. Five other observatories in Leiden, Nice, Paris, San Fernando and Washington DC were subsequently asked and it turned out that they all had a large archive, see below. The nine observatories without archives were, roughly from south to north, those in Bonn, Göttingen, Hamburg, Kiel, Aarhus, Lund, Oslo, Stockholm, and Uppsala. I therefore informed my correspondents from these nine observatories about this result asking whether they need to modify their negative answers about an archive. Uppsala, Göttingen, Bonn, Aarhus and Lund replied as included below.

Finding that nine out of the first ten observatories inquired had no archive at all from those years was such a surprise that I asked the question several times during half a year to the more than twenty persons on my list, always with the same negative result. For example, Michael Geffert wrote from Bonn "Unfortunately, we don't have a real archive."

An explanation was given by one of my correspondents who wants to remain anonymous: "The preservation in archives is always largely a matter and responsibility of the directors of the institutes, who normally do not have an education in history, law etc. as they should have. ... I have also discarded most of my ... documents because their original purpose has been fulfilled."

An illuminating answer was given by Axel Wittmann, retired member of the Göttingen University Observatory (USG), when he told what happened to the archive of Otto Hechmann. Heckmann seemed especially interesting because of his initiative to arrange the Baltic meetings (Høg 2015a) and for his great significance for astronomy all together.

Wittmann wrote: "Heckmann was a member of the "Gauss Society" of which I am the secretary since 2002. Today I am taking care of the Gauss Society's archive at the USG (which is now called "Historische Sternwarte" and - comfortably refurbished - houses the Lichtenberg-Kolleg, a new institute founded in ca. 2008)."

"When Otto Heckmann had died his personal belongings went to the family. As far as I remember only some of his professional books were sold to interested colleagues and to professional antiquarians: I was able (and could afford...) to buy a few of his books (2 or 4 if I remember correctly) from the family through Professor Voigt. There is no Heckmann archive at Göttingen, neither in the USG nor in the IAG or in the University Library (SUB). The only 'archive' which has come down to us (viz. to the Gauss Society) is a collection of reprints about cosmology, which I have "preserved" from being thrown away until today; and this is a property of the Gauss Society of which Otto Heckman was a member. Many years ago I had asked our University Library whether they would like to get and to preserve it. The answer was that "we do not collect offprints - we have the originals" (in the various journals, I guess....). "

In order to investigate whether such a small fraction of observatories kept archives as I had found (one out of ten) I wrote in May 2015 to colleagues at the observatories in San Fernando, Washington DC, Paris, Nice and Leiden: "Does your observatory have an archive of letters, administrative and scientific? Would you please tell me approximately how many letters it contains for the 1950s and 60s? I am collecting such information from many observatories with quite surprizing results."

All five observatories answered that in fact they had archives and their answers follow here.

**San Fernando**
Answer on 28 May from Javier Montojo Salazar:
Yes, of course we have a quite large historical archive (AHROA) since the creation of the "Real Observatorio", first in Cádiz (1753) and latter in San Fernando (1798). The number of documents dated in the 50s and 60s are around 20,000.

Best regards,
Javier

**US Naval Observatory**
Answer on 29 May from Gregory Shelton:

**Subject: RE: Archive of US Naval Observatory Letters**
The Naval Observatory's archives are in Records Group 78 at the United States National Archives, http://research.archives.gov/description/407

The papers of Kaj Aage Gunnar Strand, astronomer, writer and lecturer were given to the Naval Observatory Library by Dr. C. Vibeke Strand, daughter of Kaj and Emilie Rashevsky Strand in 2003. Processing history as related by Brenda Corbin and Jody Armstrong. Papers were later deposited in the Library of Congress Manuscript Division.

Dr. Steven Dick's book Sky and Ocean Joined (2003) has additional indexes and footnotes to archives related to the Naval Observatory.

Hope this helps, Gregory

James Melville Gilliss Library
United States Naval Observatory
3450 Massachusetts Avenue, N.W.
Washington, D.C. 20392-5420
Phone: 202-762-1463
e-mail: gregory.shelton@navy.mil
web site: http://www.usno.navy.mil/USNO/library

**Paris Observatory**
Answer on 1 June from an anonymous librarian:

Dear Mr Hog,

It's difficult to give you a number, even approximate, of administrive and scientific letters in our archive. Indeed, the letters are classified by name or thematic such as comptability, equipment, Observatory staff, etc ...., wherein you have letters.
However, we have two archive boxes of administrative and scientific letters of André Danjon (1946-1959).
You can follow this link to see the inventory of our administrative and
scientific correspondances :

https://pleade.obspm.fr/sdxapp/results.html?base=ead&champ1=fulltext&op1=AND&search_type=simple&query1=correspondance*+administrative*&ssearch-submit-npt.x=7&ssearch-submit-npt.y=6

Best regards,
--
La.Bibliothèque
Service de questions-réponses de la Bibliothèque de l'Observatoire de Paris
http://www.obspm.fr/bibliotheque.html

Bibliothèque
Section de Paris
Adresse postale :
61 Avenue de l'Observatoire
75014 PARIS
Accès visiteurs :
77 Avenue Denfert-Rochereau

Bâtiment Perrault
+33 1 40 51 21 41
+33 1 40 51 21 90
http://www.obspm.fr/acces-au-site-de-paris.html

Bibliothèque
Section de Meudon
Bâtiment Evry Schatzman n°18 (LAM)
5 place Jules Janssen
92195 MEUDON
+33 1 45 07 79 34
http://www.obspm.fr/acces-au-site-de-meudon.html

### Leiden Observatory

> Answer on 23 June from Jan Lub:
> Beste Erik,

Re: Archive.
1. Yes we have
2. The papers of J.H. Oort are preserved in the University Library.
Jet Katgert has produced a book with a description of this archive in 1997.
The Letters and Papers of Jan Hendrik Oort ISBN 0-7923-4542-8 published by Kluwer.
3. We have collections also from Henk van de Hulst and (partially) from Mayo Greenberg and Harry van der Laan and most likely also from Willem de Sitter and Ejnar Hertszprung, but no system.
4. Papers from Adriaan Blaauw are if any in Groningen and at ESO.

Jan

### Nice Observatory

```
Answer on 17 July from Christine Etienne:

Responsable du Service Patrimoine
Observatoire de la Côte d'Azur
Boulevard de l'Observatoire
CS 34229
06304 Nice Cedex 4
```

# Archives du Service Patrimoine de l'Observatoire de la Côte d'Azur pour les années 1950 et 1960

## Archives administratives

**« Vie de l'Observatoire 1941-1961 »** : correspondances, devis, factures (concernant l'eau, électricité, téléphone, budget marchés, dépenses, véhicules, entretiens, petits matériels, travaux, personnels (promotions, nominations, salaires, indemnités, bourses, allocations, mobilisation, retraites, titularisation)).
Une boîte « Cauchard » complète (de 34 x 27 x 10 cm) pour la période 1941-1961

## Archives administratives et scientifiques de Bernard Milet (1959/1982)
Arrêtés nomination, primes, comptes rendus des observatoires. Courriers The Cincinnati Observatory – Paul Herget : courriers 1963, 1966, 1967, 1968. Notes relatives à la protection de l'observatoire de Haute-Provence contre les lumières, les fumées, les poussières etc., 17 décembre 1959.
Plusieurs dizaines de pièces.

## Archives scientifiques et administratives de Michel Hénon
Approximativement 5 boites à archives classiques (ces archives s'étendent beaucoup plus que les années 1960 donc cette estimation est approximative).

## Archives scientifiques

**Quelques aérogrammes de Cincinnati, Ohio 1953/1957.**

**Fiches d'observation relatives aux plaques photographiques de l'astrographe de Nice**
1068 fiches d'observation de Bernard Milet de 1965 à 1969
524 fiches d'observation de Marguerite Laugier de 1950 à 1961

**Copies de cahiers d'observations du télescope de Schmidt de Meudon** : 3 cahiers :
1) 137 pages de 1961 à 1964
2) 111 pages de 1964 à 1966
3) 100 pages de 1966 à 1969

**Cahiers d'observations de la Coupole Charlois (38 cm et 52 cm) de Nice** : 3 cahiers (1964 à 1968)

**Une partie de cahier d'observations de la Grande coupole (76 cm) de Nice**.
L'ensemble du cahier recouvre la période 1969 à 1978.

---

**Uppsala Observatory**
From Bengt Edvardsson I hear in August that they will take a new look at the question of archiving in view of my report. - No later information has been received.

### Göttingen Institute for Astronomy

Wolfram Kollatschny writes in August: "It is true that the institute for astronomy at Göttingen has no archiv. However, we have collected a few older manuscripts and some instrument parts. Furthermore, the university of Göttingen has an archiv. There they have a lot of information about Gauss and Schwarzschild."

### Bonn Observatory

Michael Geffert writes in August:

Dear Erik,

please find the following information about an archive of astronomy in Bonn observatory:

The observatory of Bonn, Observatorium Hoher List, was closed in the middle of 2013. All of the historical material was taken to the Argelander-institute in Bonn. We don't have an own archive in this institute. Historically interesting letters and other writings (e.g. from Argelander, Schönfeld, Küstner) were collected in the main archive of university. As far as I know, no letters from the time after the second world war were collected.

On the other hand, the institute is establishing the "Sammlung Historischer Himmelsaufnahmen" (https://astro.uni-bonn.de/de/offentlichkeit/sammlung-historischer-himmelsaufnahmen), where historical telescopes, books and mainly the photographic observations of the observatory werde collected. With special projects, exhibitions and talks for the public, we want to bring the history of astronomy in Bonn to the attention of many people. More informations may be found in short videos (in German) at Youtube ("Sammlung Historischer Himmelsaufnahmen").

Best wishes

Michael

### Aarhus Observatory

Palle Lykke, historian of the Aarhus University, informs me in August that no archive exists from Mogens Rudkjøbing, professor of astronomy in those years,

### Lund Observatory

Knut Lundmark (1889-1958), director of Lund Observatory, worked for 40 years on a catalog of galaxies, the "Lund General Catalog" , but it was never published. The book about Lundmark by Sundman (1988) notes on p. 140 that the cards for the catalog have disappeared.

When Lennart Lindegren was asked in November he answered:

The letters of Knut Lundmark are at the Lund University Library (not at
the Observatory). I think it is basically correct that there are no
archives at the Observatory from the 1950s.

However, the Lund General Catalogue (or large parts of it at least) still
exists and is kept in our basement.

**Soon after I saw the cards for the catalog there stored in a two meter high top-filled bookshelf.**